\documentclass[12pt]{iopart}

\usepackage{iopams}  
\usepackage{graphicx}
\usepackage{bm}

\begin{document}

\title[Quantum tunneling of composite object coupled with quantized radiation field]{Quantum tunneling of composite object coupled with quantized radiation field}

\author{Fumika Suzuki}

\address{Department of Physics \& Department of Chemistry, University of British Columbia, Vancouver, British Columbia, Canada V6T 1Z1}
\address{Physikalisches Institut, Albert-Ludwigs-Universit\"{a}t Freiburg, 79104 Freiburg, Germany}

\ead{fumika@physics.ubc.ca}

\author{Friedemann Queisser}
 \address{%
Fakult\"{a}t f\"{u}r Physik, Universit\"{a}t Duisburg-Essen, Lotharstrasse 1, 47057 Duisburg, Germany
}
\vspace{10pt}

\begin{abstract}
We study quantum tunneling of a composite object, which has a dipole or quadrupole moment  coupled with  quantized (photon or gravitational) radiation field, through a $\delta$ potential barrier. The dipole or quadrupole moment    is represented by 
an oscillator in the relative coordinate of two  constituent particles of the object. The center of mass degrees of freedom of the object is not directly coupled with the radiation field. However, we show that, for the object with the oscillator  in the excited  state, dissipation caused by the radiation field can suppress its quantum tunneling rate in the center of mass degrees of freedom. In addition, when the initial energy of its center of mass motion is similar to that of the excited state of the oscillator, a spatial superposition state of the object prepared by the  barrier  can decohere due to the radiation field. The main purpose of this article is to investigate how two different interplays (i) among the center of mass, the relative coordinate degrees of freedom and the potential barrier, and (ii) between the relative coordinate degrees of freedom and the radiation field, can affect the quantum tunneling and the creation of the spatial superposition state of the  object. Our toy model can give insight into tests of quantum tunneling and quantum superposition of atoms or molecules with its dipole or quadrupole moment coupled with the radiation field.
\end{abstract}

\section{Introduction}

The study of quantum phenomena of composite objects  has various applications in diverse fields, such as investigations into quantum-to-classical transition \cite{arn}, in the study of effects of gravity and relativity on quantum mechanics \cite{pik, cisco}, in quantum chemistry \cite{taka} and in quantum computing \cite{taka2}. 

Recently quantum tunneling of a composite object through a $\delta$ potential barrier has been studied in continuous space \cite{que} and on a lattice \cite{fumika2}. It was found that its internal structure can produce phenomena such as long-lived resonances and bound states in the continuum that can not be seen in a single particle quantum tunneling. 

In this article, we investigate a quantum tunneling of a composite object by using the model which was introduced in \cite{que}, but now the object is coupled with  quantized radiation field due to radiation emission from its dipole or quadrupole moment. It is then observed that although dissipation and decoherence due to the radiation field act only on one degrees of freedom (i.e., relative coordinate degrees of freedom of the constituent particles of the object), yet have observable consequences for the other degrees of freedom (i.e., the center of mass (c.m.) degrees of freedom of the object) when those two degrees of freedom are correlated by, e.g., propagation through the potential
barrier. This kind of effects can be seen in many situations in physics, though their theoretical treatments have not been well investigated. Here we  see the effect
occurring in a relatively simple model.  

Decoherence due to dipole or quadrupole moment radiation has been studied in the literature \cite{bre, bre2, fumika} in the context of quantum field theory. In the literature, dipole moment or quadrupole moment of a composite system consisting of a particle and device (e.g., slit, mirror) which recombines paths of a particle was considered. In such a situation, cutoff of  radiation spectrum and decoherence rate should be largely depend on how the device recombines those paths and one needs to find relevant parameters of devices to estimate actual decoherence rate. On the other hand, dipole or quadrupole moments of atoms and molecules can be found easily. This is another reason why we study a superposition state of the composite object consisting of two particles as a toy model of  atom or molecule.

The article is organized as follows. Firstly we introduce the Lagrangian for the composite object whose dipole moment described by an oscillator in the relative coordinate of two constituent particles is coupled with the photon radiation field. Then we derive the master equation for the  object by tracing out the radiation field (section \ref{photon}). In section \ref{photon2}, we use the master equation to study its quantum tunneling through a $\delta$ potential barrier. When the object  initially has the oscillator  in the first excited state, we find that dissipation caused by the radiation emission can suppress its quantum tunneling rate in the c.m. degrees of freedom. Furthermore, when the initial energy of the c.m. motion is similar to that of the excited states of the oscillator, a spatial superposition state of the object created by the barrier (i.e., being reflected or transmitted by the barrier in the c.m. degrees of freedom) can decohere by the radiation emission (section 2.3). The c.m. degrees of freedom of the object is not directly entangled with the radiation field. However the decoherence in the degrees of freedom can happen due to the fact that the interaction with the potential barrier can cause the entanglement between the c.m. and the relative coordinate degrees of freedom which is entangled with the radiation field. Therefore, the main purpose of our study is to investigate how two different interplays (i) among the c.m., the relative coordinate degrees of freedom and the potential barrier, and (ii) between the relative coordinate degrees of freedom and the radiation field,  can affect the quantum tunneling rate and creation of the spatial superposition state of the composite object by the barrier.

This article deals largely with the photon radiation field. However, since the studies of gravitational decoherence have attracted increasing interest recently \cite{penrose, diosi, ble, cisco2, mari} and electromagnetism and linearized gravity can be treated similarly, we provide  section \ref{gravity} which addresses the case where the quadrupole moment of the object is coupled with the quantized gravitational radiation field by emission of gravitational waves. Finally we briefly discuss the case where one introduces the Lorenz gauge  and the false loss of coherence observed in the gauge, and the problem of quantum gravity which arises when the field is replaced by gravity in section \ref{lorenzgauge}.

 Units are chosen throughout such that $c=\hbar=1$.
 
 \section{Composite object coupled with  the quantized photon  field}

In this section, we derive the master equation for the composite object whose dipole moment is coupled with the photon radiation field. The c.m. degrees of freedom of the object is not directly coupled to the radiation field, while it gets entangled with the relative coordinate degrees of freedom of its constituent particles by propagation through the barrier. Using the master equation, we find that dissipation and decoherence caused by the radiation field affect the quantum behavior of the object in the c.m. degrees of freedom.

\subsection{The master equation}\label{photon}

We derive the master equation for the composite object coupled with the photon radiation field using the path integrals formalism. We consider the Lagrangian for the  object which consists  of two particles interacting with each other by a harmonic potential:
\begin{eqnarray}\label{composite}
L_0=\frac{1}{2}M\mathbf{\dot{X}}^2+\frac{1}{2}\mu \dot{\mathbf{x}}^2-\frac12 \mu \omega^2 \mathbf{x}^2
\end{eqnarray}
where $\mathbf{X}$ is the position of c.m. for two particles at $\mathbf{x}_1$ and $\mathbf{x}_2$, $\mathbf{x}$ is the relative distance between them, $M$ is the sum of two masses $M=m_1+m_2$ and $\mu$ is the reduced mass. 

The dipole moment of the object is represented by an oscillator in the relative coordinate $\mathbf{x}$ of the two constituent particles, and  $\omega$ is the natural frequency of the  oscillator. In this article, the two particles bounded by the harmonic potential is a toy model for ``the composite object" and its position is given by the c.m. position $\mathbf{X}$. We study its quantum tunneling through the barrier $V (\mathbf{x},\mathbf{X})$ while its dipole moment is 
coupled with the quantized radiation field by emitting radiation.

Firstly we derive the influence functional for  coupling to the photon radiation field. Here we work in the Coulomb gauge, $\nabla \cdot \mathbf{A}=0$ . We express the vector potential $\mathbf{A}(\mathbf{r},t)$ and the current density $\mathbf{j}(r,t)$ in terms of the Fourier transform of the dynamical variables
\begin{eqnarray}
\mathbf{A}(\mathbf{r},t)=\sum_{\mathbf{k}} \mathbf{a}_{\mathbf{k}}(t) e^{i\mathbf{k}\cdot \mathbf{r}},\quad
\mathbf{j} (\mathbf{r},t)=\sum_{\mathbf{k}} \mathbf{j}_{\mathbf{k}} (t) e^{i\mathbf{k}\cdot \mathbf{r}}.
\end{eqnarray}

Assuming two particles have the opposite charges $q$ and $-q$, the action for the  object coupled to the radiation field can be written as  \cite{feynman}
\begin{eqnarray}
S=S_V +S_{\rm int}+ S_{\rm rad}
\end{eqnarray}
where
\begin{eqnarray}\label{action}
S_V&=&\int (L_0-V(\mathbf{x},\mathbf{X})) dt,\nonumber\\
S_{\rm int} &=& \int\int d^3 \mathbf{r} dt \mathbf{j} (\mathbf{r},t) \cdot \mathbf{A} (\mathbf{r},t)=q \int \displaystyle\sum_{\mathbf{k},\lambda=1,2}\dot{x}_{\lambda}a_{\lambda,\mathbf{k}}e^{i\mathbf{k}\cdot \mathbf{x}_{\mathcal{O}}(t) }dt,\nonumber\\
S_{\rm rad}&=&-\frac14 \int d^4 r F^{\mu\nu}F_{\mu\nu}=\frac12 \int \displaystyle\sum_{\mathbf{k},\lambda=1,2}(\dot{a}_{\lambda ,\mathbf{k}}^{*}\dot{a}_{\lambda, \mathbf{k}}-k^2 a_{\lambda, \mathbf{k}}^{*}a_{\lambda,\mathbf{k}}) dt.
\end{eqnarray}
Here we used the fact that the current density of two charges $\mathbf{j}(\mathbf{r},t)=q\dot{\mathbf{x}}_1 \delta (\mathbf{r}-\mathbf{x}_1(t))-q\dot{\mathbf{x}}_2\delta (\mathbf{r}-\mathbf{x}_2(t))$ can be approximated as an oscillating dipole with origin at the center of the charge distribution $\mathbf{x}_{\mathcal{O}}$, $\mathbf{j}(\mathbf{r},t)=\dot{\mathbf{\chi}}\delta (\mathbf{r}-\mathbf{x}_{\mathcal{O}}(t))$ and the corresponding $\mathbf{j}_{\mathbf{k}}(t)=\dot{\chi}e^{-i\mathbf{k}\cdot\mathbf{x}_{\mathcal{O}}(t)}$ where $\chi (t)=q\mathbf{x}(t)$ if we keep only the lowest-order term in a Taylor series with $\mathbf{x}_{\mathcal{O}}$ \cite{nano}. $\lambda=1,2$ represents polarizations and $a_{1,\mathbf{k}}$ and $a_{2,\mathbf{k}}$ are the components of $\mathbf{a}_{\mathbf{k}}$ in two directions perpendicular to $\mathbf{k}$. Similarly $x_{1}$ and $x_2$ are the components of $\mathbf{x}$ in the direction transverse to $\mathbf{k}$. For each $\lambda$, we have Maxwell's equations, $\ddot{a}_{\lambda,\mathbf{k}}+k^2 a_{\lambda,\mathbf{k}}=j_{\lambda,\mathbf{k}}$ \cite{feynman}.

Now the reduced density matrix $\rho(\mathbf{X},\mathbf{X}',\mathbf{x},\mathbf{x}',t)$ for the  object can be given by tracing out the photon radiation field \cite{feynman2},
\begin{eqnarray}\label{reducedint}
&&\rho(\mathbf{X},\mathbf{X}',\mathbf{x},\mathbf{x}',t)=\int d\mathbf{X}_0d\mathbf{X}'_0 d\mathbf{x}_0 d\mathbf{x}_0' \int\int\int\int \mathcal{D}\mathbf{X}\mathcal{D}\mathbf{X}'\mathcal{D}\mathbf{x}\mathcal{D}\mathbf{x}'\mathcal{F} (\mathbf{j},\mathbf{j}')\nonumber\\
&&\qquad\qquad\times e^{iS_V(\mathbf{X},\mathbf{x})-iS_V(\mathbf{X}',\mathbf{x}')}\rho (\mathbf{X}_0,\mathbf{X}'_0,\mathbf{x}_0,\mathbf{x}_0',0)
\end{eqnarray}
where $\mathbf{j}$ and $\mathbf{j}'$ represent current associated with $\mathbf{x}$ and $\mathbf{x}'$ respectively. Since $S_{\rm rad}+S_{\rm int}$ gives the action for the forced harmonic oscillator, we can derive the influence functional $\mathcal{F} (\mathbf{j},\mathbf{j}')$ by assuming that the initial wavefunctional for the photon radiation field is in the ground state:
\begin{eqnarray}\label{influence}
\mathcal{F} (\mathbf{j},\mathbf{j}')=\exp i\Theta_{\rm photon}\nonumber\\
=\exp \Biggr[i q^2\displaystyle\sum_{\mathbf{k}}\int_0^{t}\int_0^{s}ds ds' (\dot{x}_{i}(s)e^{i\mathbf{k}\cdot \mathbf{x}_{\mathcal{O}}(s)}-\dot{x}'_{i}(s)e^{i\mathbf{k}\cdot \mathbf{x}'_{\mathcal{O}}(s)})\gamma^{ij}(s,s')\nonumber\\
\qquad\qquad\qquad\qquad\qquad \times(\dot{x}_{j}(s')e^{-i\mathbf{k}\cdot \mathbf{x}_{\mathcal{O}}(s')}+\dot{x}'_{j}(s')e^{-i\mathbf{k}\cdot \mathbf{x}'_{\mathcal{O}}(s')})\nonumber\\
-q^2\displaystyle\sum_{\mathbf{k}}\int_0^{t}\int_0^{s}ds ds' (\dot{x}_{i}(s)e^{i\mathbf{k}\cdot \mathbf{x}_{\mathcal{O}}(s)}-\dot{x}'_{i}(s)e^{i\mathbf{k}\cdot \mathbf{x}'_{\mathcal{O}}(s)})\eta^{ij}(s,s') \nonumber\\
\qquad\qquad\qquad\qquad\qquad \times (\dot{x}_{j}(s')e^{-i\mathbf{k}\cdot \mathbf{x}_{\mathcal{O}}(s')}-\dot{x}'_{j}(s')e^{-i\mathbf{k}\cdot \mathbf{x}'_{\mathcal{O}}(s')})\Biggr]
\end{eqnarray}
where the correlation functions are given by summing over photon polarizations,
\begin{eqnarray}
\gamma^{ij}(s,s')&=&\delta^{ij}\frac{1}{2k}\sin k (s-s'), \quad \eta^{ij} (s,s')=\delta^{ij}\frac{1}{2k}\cos k (s-s')
\end{eqnarray}
with $i, j=1,2$ associated to two directions transverse to $\mathbf{k}$. 

We now derive the master equation. The relative coordinate degrees of freedom $\mathbf{x}$ of the  object follows an equation of motion for a harmonic oscillator $L_0$ (\ref{composite}). For simplicity, we assume that the object moves in a one-dimensional space. Ignoring damping of the 
motion in the relative coordinate due to radiation, the classical path $x^{c}(s)$ with the boundary condition $x(0)=x_0$, $x(t)=x$ can be written as
\begin{eqnarray}\label{semiclassical}
x^{c}(s)=x_0\frac{\sin \omega (t-s)}{\sin \omega t}+x\frac{\sin \omega s}{\sin\omega t}.
\end{eqnarray}

We substitute (\ref{semiclassical}) into (\ref{influence}) to obtain
\begin{eqnarray}\label{Theta}
\mbox{Re}(\Theta_{\rm photon})&\approx &\frac{q^2\omega}{4\pi^2}\int_0^{t} (\dot{x} (s)^2-\dot{x}' (s)^2) ds\nonumber\\
&+&\frac{q^2\omega^2}{8\pi} \int_0^{t}(\dot{x} (s)-\dot{x}'(s))(x(s)+x'(s))ds,\nonumber\\
\mbox{Im}(\Theta_{\rm photon})&\approx &\frac{q^2\omega^2}{8\pi^2}\int_0^{t} (\dot{x}(s)-\dot{x}'(s)) (x (s)-x'(s))ds\nonumber\\
&\approx& \frac{q^2\omega^2}{16\pi^2} (x(t)-x'(t))^2 
\end{eqnarray}
where  it was estimated that a natural upper cutoff for the frequency spectrum of the emitted radiation is  the natural frequency $\omega$ of the  oscillator and $\omega t\gg 1$. We also  introduced the electric dipole approximation where $e^{-i\mathbf{k}\cdot \mathbf{x}_{\mathcal{O}}(s')}$ etc. can be approximated by unity.   

Then the density matrix for the  object $\rho (X, X', x, x', t)$ at time $t$ is given by
\begin{eqnarray}\label{density}
&&\rho (X, X', x, x', t)\nonumber\\
&&=\int\int d X_0 d X'_0 d x_0 dx'_0 W (X, X',x,x',t;X_0, X_0', x_0, x'_0, 0)\nonumber\\
&&\qquad\qquad\qquad\qquad\qquad\qquad\times \rho (X_0, X'_0, x_0, x'_0, 0)
\end{eqnarray}
where the reduced density matrix propagator is
\begin{eqnarray}
&&W (X, X',x,x',t;X_0, X_0', x_0, x'_0, 0)\nonumber\\
&&=\int\int\int\int\mathcal{D} X\mathcal{D} X'\mathcal{D} x\mathcal{D} x'\exp i \Biggr(S_V [x, X]-S_V [x',X']\nonumber\\
&&+\frac{q^2\omega}{4\pi^2}\int_0^{t}(\dot{x}(s)^2-\dot{x}'(s)^2)ds\nonumber\\
&&+\frac{q^2\omega^2}{8\pi}\int_0^{t}(\dot{x}(s)-\dot{x}'(s))(x(s)+x'(s))ds\Biggr)\nonumber\\
&&\times\exp \left(-\frac{q^2\omega^2}{8\pi^2}\int_0^{t}  (\dot{x} (s)-\dot{x}' (s))(x(s)-x'(s))  ds\right)
\end{eqnarray}
and
\begin{eqnarray}
S_V [x,X]=\int_0^{t}\left(\frac{M}{2}\dot{X}(s)^2+\frac{\mu}{2}\dot{x}(s)^2 -\hat{V}' (x,X) \right)ds.\nonumber\\
\end{eqnarray}

Here we rewrote the potential term $V' (x,X)=\frac12 \mu \omega^2 x^2+V(x,X)$.

Differentiating both sides of (\ref{density}) with respect to $t$, we obtain the master equation \cite{cl}:
\begin{eqnarray}\label{master}
\frac{\partial \rho (X, X', x, x', t)}{\partial t}\approx\Biggr[\frac{i}{2M}\left(\frac{\partial^2}{\partial X^2}-\frac{\partial^2}{\partial X'^2}\right)+\frac{i}{2\mu+q^2\omega/\pi^2}\left(\frac{\partial^2}{\partial x^2}-\frac{\partial^2}{\partial x'^2}\right)\nonumber\\
-i (V'(x,X)-V'(x',X'))+\frac{ q^2 \omega^2  }{8\pi \mu+4q^2\omega/\pi}(x+x')\left(\frac{\partial}{\partial x}
+\frac{\partial }{\partial x'}\right)\nonumber\\
-\frac{q^2 \omega^2 }{8\pi^2 \mu+4q^2\omega}(x-x')  \left(-i\frac{\partial}{\partial x}-i\frac{\partial}{\partial x'}\right) \Biggr]\rho (X,X', x, x', t).\nonumber\\
\end{eqnarray}

We use the above master equation in the following sections.

\subsection{Quantum tunneling and dissipation}\label{photon2}

\begin{figure}
\begin{center}
{%
\includegraphics[clip,width=1\columnwidth]{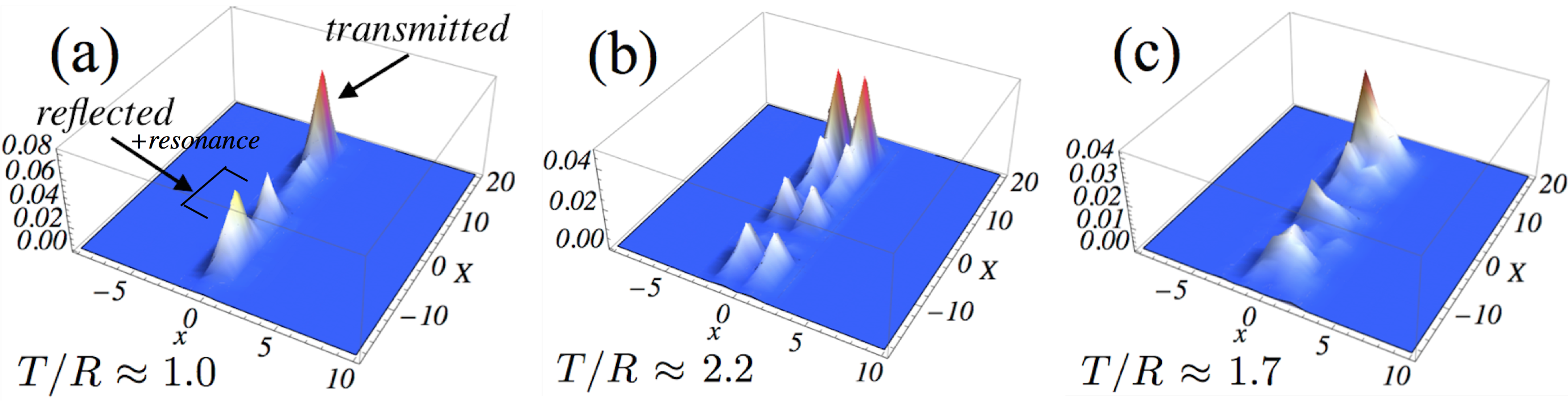}%
}
\caption{The probability distribution $\rho (X,X,x,x,t')$ of the composite object after  tunneling through the  barrier. $m_1=m_2=1$, $\omega t'=11\times 10^{9}$, $(P_0^2/2M)/V_1=(P_0^2/2M)/V_2=1.2$ and $\sigma/x_{a,0}=6.6$. (a) The oscillator in the relative coordinate is initially in the ground state (without dissipation). (b)  The oscillator in the relative coordinate is initially in the first excited state (without dissipation). (c)  The oscillator in the relative coordinate is initially in the first excited state (with dissipation).}
\end{center}
\label{diss}
\end{figure}

Spontaneous emission from two-level atoms tunneling through a square potential and a double-well potential has been studied in the literature \cite{jap, bra}. In some cases, dissipation can enhance quantum tunneling effects due to  increase of kinetic energy \cite{jap} or  mixing of excited states with  ground states caused by coupling with  environment \cite{sasaki}. However it is known that dissipation generally suppresses quantum tunneling effects \cite{cl2}. Here we study a quantum tunnelling of the composite object through a $\delta$ potential barrier, $V(x, X)=V_1 \delta (x_1)+V_2\delta (x_2)=V_1 \delta (X+\frac{m_2}{M}x)+V_2\delta (X-\frac{m_1}{M}x)$, while its dipole moment radiates photons. We assume $m_1=m_2=1$, $V_1=V_2$ and $q$ is equal to an elementary charge $e$. For simplicity, we introduce rescaling of the coordinate $X\rightarrow 2X=x_1+x_2$, while $x=x_1-x_2$. The two constituent particles of the composite object move only in one dimension and the line joining them lies along the dimension perpendicular to the plane of the barrier oriented along the other two dimensions in three-dimensional space. We start with the initial wave packet of the form $\Psi_0 (X, x)=f(X)\phi_{n} (x)$ at $t=0$:
\begin{eqnarray}\label{wave}
f(X) &=& \left(\frac{2}{\pi\sigma^2}\right)^{1/4}\exp \left(i P_0 X-\frac{(X-X_{\mathcal{O}})^2}{\sigma^2}\right),\nonumber\\
\phi_{n} (x)&=&\frac{1}{\sqrt{2^{n} n!}}\left(\frac{\mu \omega}{\pi}\right)^{1/4} e^{-\mu \omega x^2/2} H_{n} (\sqrt{\mu\omega} x)
\end{eqnarray}
 where $P_0$ is the initial c.m. momentum, and $\sigma$ is the width of the initial wave packet with its center being $X_{\mathcal{O}}$.

Here $X_{\mathcal{O}}=-10$ so that it tunnels through the barrier located around $X\sim 0$ from the left. We choose the ratio of  kinetic energy of the c.m. motion divided by the strength of the
potential barrier $(P_0^2/2M)/V_1=(P_0^2/2M)/V_2=1.2$, and $\sigma/x_{a,0}=6.6$ where $x_{a,n}$ is  the oscillator amplitude given by $x_{a,n}=\sqrt{2 E_{n}/\mu\omega^2}$ with $E_{n}=(n+1/2)\omega$. We let the kinetic energy of the c.m. motion is much less than the energy of the first excited state of the oscillator in the relative coordinate, $\frac{3\omega/2}{P_0^2/2M}=23$. We define the transmission probability $T=\int_{X>0}^{\infty}dX\int_{-\infty}^{\infty}dx\rho (X,X,x,x,t') $ and the reflection+resonance probability $R=\int_{-\infty}^{X\leq 0}dX\int_{-\infty}^{\infty}dx\rho (X,X,x,x,t') $ where $t=t'=3$ is the time when we stop the time evolution of the wave packet after it has been well-separated into the transmitted, resonance and reflected parts by the interaction with the barrier. Note that resonances at the barrier found in \cite{que} are also observed  and their contributions are included in the reflection+resonance probability $R$ here instead of waiting for their complete decay after a long time evolution to reduce numerical evaluations.

When the object is not coupled with the radiation field, it is found that $T/R\approx 1.0$ (i.e., half of the wave packet is transmitted and half
of it is reflected) if its  oscillator  is initially in the ground state (Fugure 1 (a)), while $T/R\approx 2.2$ if the oscillator is initially in the first excited state (Figure 1 (b)). Here $\omega t' =11\times 10^9$ is chosen. This shows that the transmission probability of the object with the excited  oscillator is around twice larger than that of the object whose oscillator is initially in the ground state. It is because the wave packet is populated around $x =0$ when the  oscillator is in the ground state and it feels $V = (V_1 +V_2)\delta (X)$ which is generally a higher potential barrier than that at $x \not= 0$. This occurs when the barrier width is narrower than the width of the ground state wave-function of the  oscillator. Because
the barrier here is a delta function of infinitesimal width, it always
occurs in this model.

\begin{figure}
\begin{center}
{%
\includegraphics[clip,width=0.5\columnwidth]{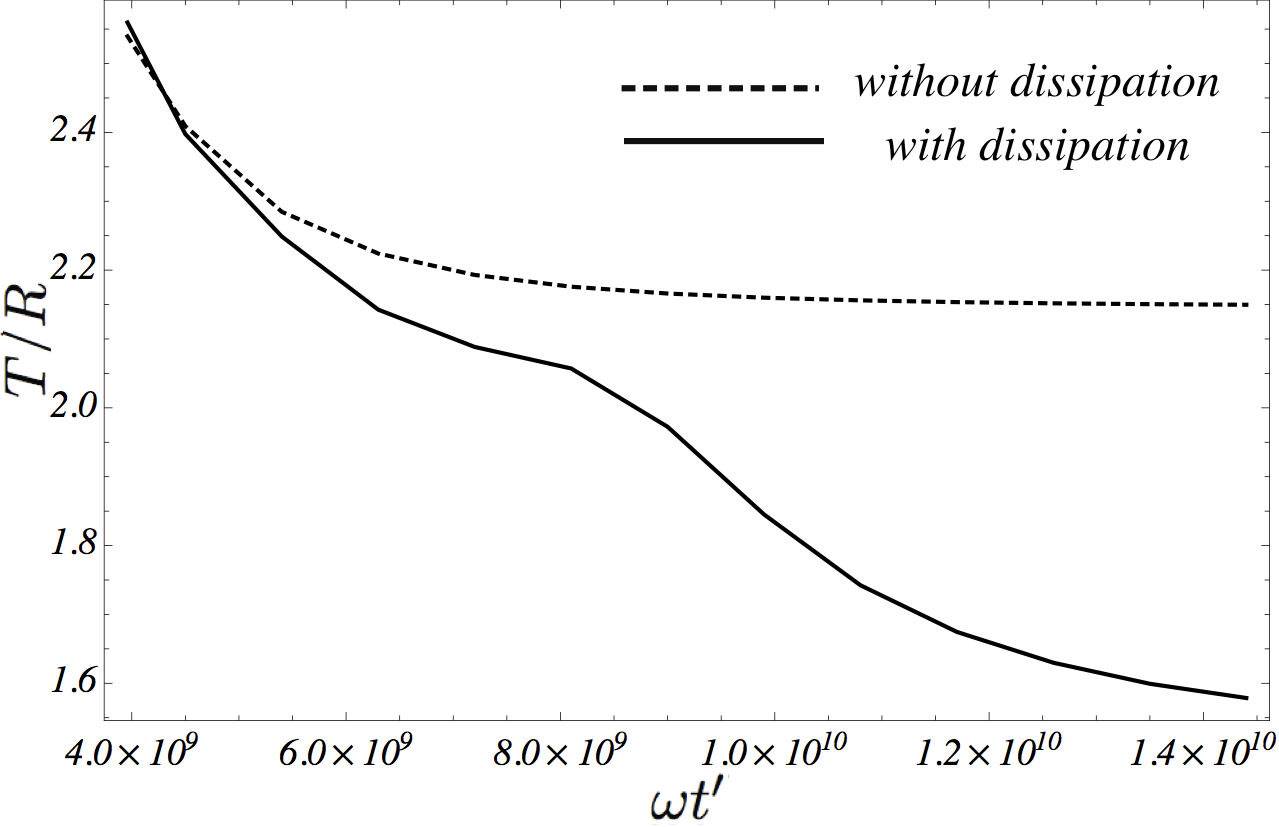}%
}
\end{center}
\caption{$T/R$ with different $\omega t'$ where the  oscillator is prepared initially in the first excited state according to the master equation (\ref{master}). $t'$ is fixed, and $\omega$ is increased.}
\label{displot}
\end{figure}

Now we consider a situation where the object is coupled with the radiation field. If there exist ambient photons, its quantum tunneling can be enhanced or assisted since interactions with them can excite the  oscillator. However if we have more photon emissions from the  oscillator, then generally dissipation caused by the process makes the excited states of the oscillator decay into the ground state and suppresses the quantum tunneling rate. Figure 1 (c) shows the quantum tunneling of the object coupled with the radiation field following the master equation (\ref{master}) with the same initial wave packet as that of Figure 1 (b). We see that the first excited state of the oscillator decays into the ground state and the tunnelling rate is suppressed as $T/R \approx 1.7$.  In Figure \ref{displot},  $T/R$ with different $\omega t'$ is plotted by changing $\omega$ without changing $t'=3$ where the  oscillator of the object  is prepared initially in the first excited state. The plot is obtained from the time evolution of the wave packet according to the master equation (\ref{master}). Without dissipation due to  the radiation field, it is observed that $T/R$ changes only slightly with $\omega t'$. Note that $T/R$ decreases even without dissipation since $x_{a,1}$ becomes smaller as $\omega $ increases and the wave packet becomes more populated around $x=0$. However with dissipation, we see that the tunneling rate gets more suppressed as $\omega t'$ increases. This indicates that the probability of the  oscillator having
decayed into the ground state increases with $\omega t'$. Then the tunneling rate in the c.m. degree of freedom becomes suppressed since the wave packet sees a high potential barrier when its  oscillator is in the ground state. 

\subsection{Decoherence}

In addition to the dissipation effects discussed in the previous section, we also observe decoherence effects. The effects can be observed when the  oscillator is prepared in a superposition state. As an example of preparation of such state, let us consider the initial wave packet of the composite object of the form (\ref{wave}) whose  oscillator is initially in the ground state and let it tunnels through the $\delta$ potential barrier as before. Similarly to the previous section, if we choose $m_1=m_2=1$, $\omega t'=9\times 10^7$ where $t'=0.03$, $(P_0^2/2M)/V_1=(P_0^2/2M)/V_2=1.2$, $\sigma/x_{a,0}=2$, half of the wave packet is transmitted and half
of it is reflected by the barrier. In the previous section, we had the energy of the c.m. motion was much less than the energy of the excited  oscillator. Here we let the energy of the c.m. motion is the same as the energy of the forth excited state of the  oscillator, $\frac{9\omega/2}{P_0^2/2M}=1$. Then the energy transfer from the c.m. motion to the  oscillator occurs when the wave packet is reflected by the barrier, and the  oscillator  gets excited partially in the reflected part, while it is still in the ground state in the transmitted part as shown in Figure 3 (a).  In Figure 3 (b),  we plot $p_{n,T}=\int_{X>0}^{\infty}dX \int dx \int dx' \phi^{*}_n (x)\rho(X,X,x,x',t')\phi_n (x') $ and $p_{n,R}=\int_{-\infty}^{X<0}dX \int dx \int dx' \phi^{*}_n (x)\rho(X,X,x,x',t')\phi_n (x') $ which are the probabilities to find the  oscillator in the state $\phi_{n}$ in the transmitted and in the reflected part respectively. Due to the symmetry of the
 barrier, the antisymmetric states are not occupied in the reflected part as discussed in \cite{que}. 

In Figure 3 (a), it can be seen that the transmitted part has the probability distribution populated around $x=0$ (i.e., the  oscillator is in the ground state) with $\rho (X=12, X'=12, x=0, x'=0,t')\approx 0.04$ while  $\rho (X=12, X'=12, x=\pm 2, x'=\pm 2,t')\approx 0$. On the other hand, the reflected part has the probability distribution populated more around $x\not= 0$ (i.e., the  oscillator is partially in the excited states) with $\rho (X=-12, X'=-12, x=\pm 2, x'=\pm 2,t')\approx 0.02$ while  $\rho (X=-12, X'=-12, x= 0, x'= 0,t')\approx 0.008$. Here we have $x_{a,0}\approx 1$, while $x_{a,2}$ is slightly larger than $2$ in the figure. 

\begin{figure}
{%
\includegraphics[clip,width=1\columnwidth]{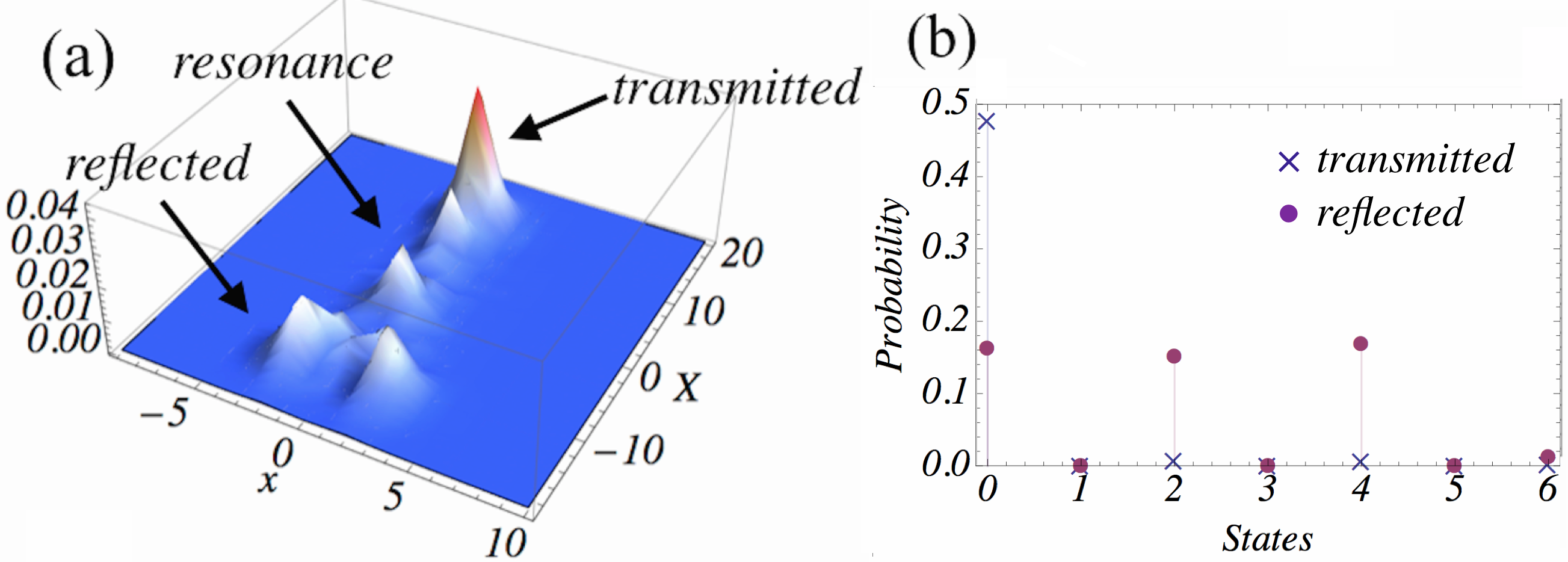}%
}
\caption{(a) The probability distribution $\rho(X,X,x,x,t')$ after tunneling through the barrier where the  oscillator was initially prepared in the ground state. The  oscillator of the reflected part gets excited partially. (b) Probabilities to find the  oscillator states $\phi_{n}$ in the transmitted and in the reflected part. State number $n = 0$ corresponds to the ground state, and $n = 1$ is for the first excited state etc. }
\label{dec}
\end{figure}

However, if we trace out the radiation field entangled with the   oscillator, as was done in the master equation (\ref{master}), then  the superposition state gets suppressed. For example, $\rho (X=12, X'=-12, x=0, x'=\pm 2,t')$ represents the superposition state of the  object being transmitted by the barrier (and arriving at $X=12$) whose oscillator is in the ground state and that being reflected by the barrier (and arriving at $X=-12$) whose  oscillator is excited. 

Since the evolution of the c.m. degrees of freedom and that of the relative coordinate degrees of freedom decouple from each other when the wave packet is far away from the barrier, we freeze the wave packet evolution in the c.m. degrees of freedom at $t=t'$ by removing the first term $\frac{i}{2M}\left(\frac{\partial^2}{\partial X^2}-\frac{\partial^2}{\partial X'^2}\right)$ of the right hand side of (\ref{master}) to simplify the numerical evaluation. Therefore the peak of the transmitted part and that of the reflected part stay at $X=12$ and $X=-12$ respectively in the discussion below.

According to the master equation (\ref{master}) without the first term involving the evolution of the c.m. degrees of freedom, we obtain
\begin{eqnarray}\label{decrate}
\mathcal{C}_{\rm dec}&=&\frac{|\rho (X=12, X'=-12, x=0, x'=\pm 2,t'+t'')|}{|\rho (X=12, X'=-12, x=0, x'=\pm 2,t')|}\approx 0.8
\end{eqnarray}
when $\omega t''= 3\times 10^9$ (i.e., $t'/t''=0.03$). Here $t''=1$ is the time measured from the moment $t=t'$. 

 Then (\ref{decrate}) shows that the superposition state of the composite object mentioned above is  suppressed $\sim 20\%$ after time $t''$.

 Note that here we have the relatively simple entanglement between the c.m. and the relative coordinate  degrees of freedom in our example, i.e., the wave-function of the composite object after the tunneling $\approx \frac{1}{\sqrt{2}}|\mbox{c.m. transmitted}\rangle|\phi_{0}\rangle +\frac{1}{\sqrt{2}}|\mbox{c.m. reflected}\rangle(\frac{1}{\sqrt{3}}|\phi_0\rangle +\frac{1}{\sqrt{3}}|\phi_2\rangle +\frac{1}{\sqrt{3}}|\phi_4\rangle)$. Then the terms in its density matrix such as $\frac{1}{2\sqrt{3}}|\mbox{c.m. transmitted}\rangle |\phi_0\rangle \langle \mbox{c.m. reflected}|\langle \phi_2|$ or $\frac{1}{2\sqrt{3}}|\mbox{c.m. transmitted}\rangle |\phi_0\rangle \langle \mbox{c.m. reflected}|\langle \phi_4|$ can be considered to represent the superposition state of the  object. However if the transmitted part also happens to have the  oscillator in the excited states, then the notion of the entanglement among the c.m., the relative coordinate  degrees of freedom and the radiation field, and the discussion of the superposition state of the  object and its decoherence can become more complicated.

\begin{figure}
\begin{center}
{%
\includegraphics[clip,width=0.5\columnwidth]{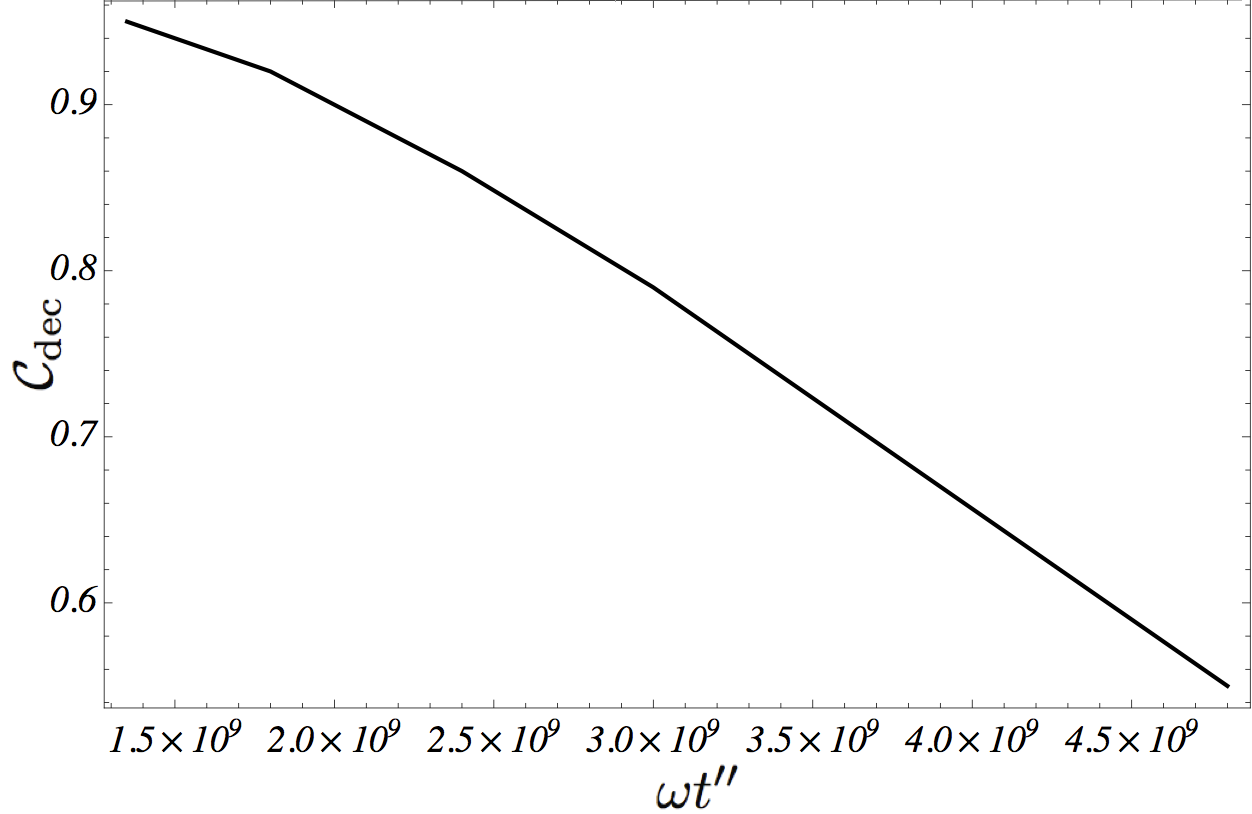}%
}
\caption{$\mathcal{C}_{\rm dec}$ with different $\omega t''$ obtained using the method in (\ref{decrate}).}
\label{decplot}
\end{center}
\end{figure}

Figure \ref{decplot} shows the dependence of $\mathcal{C}_{\rm dec}$ on $\omega t''$ obtained by changing $\omega$ without changing $t''$ by  using the master equation (\ref{master}) without the first term involving the c.m. degrees of freedom as discussed above. Apart from $\omega$, we chose same initial conditions and parameters used in the example (\ref{decrate}). The energy of the excited  oscillator and the oscillator amplitude $x_{a,n}$ depend on $\omega$. Therefore the probabilities to find the oscillator state $\phi_{n}$ in the transmitted and in the reflected part can  become slightly different from Figure \ref{dec} (b) as we change $\omega$ in Figure \ref{decplot}. However $\mathcal{C}_{\rm dec}$ purely measures the decay of the superposition state after time $t''$. As it is expected, the decoherence rate increases as $\omega t'' $ gets larger. 

In this section, we found that the interaction with the barrier can make the c.m. and the relative coordinate degrees of freedom get entangled with each other when the initial energy of the c.m. motion is similar to that of the excited states of the  oscillator. As a result, the spatial superposition state of the object gets suppressed when the radiation field is traced out even if the c.m. degrees of freedom is not directly entangled with it. On the other hand, one can avoid this decoherence effect by choosing the conditions used for Figure 1 (a). In the previous section, the initial energy of the c.m. motion was much smaller than that of the excited states of the  oscillator.  In Figure 1 (a), the energy transferred from the c.m. motion to the  oscillator during the interaction with the barrier was not enough to excite the oscillators in the reflected part, and both the reflected and transmitted parts had the same oscillator states that remained in the ground state. One can look at the time evolution of the entanglement entropy which measures the entanglement between c.m. and relative coordinate degrees of freedom, $S=-\mbox{tr}[\rho_{\rm c.m.} \log_2 \rho_{\rm c.m.}]$ where $\rho_{\rm c.m.} =\int_{-\infty}^{\infty} \rho (X,X', x,x,t) dx$ (Figure 5). During the tunnelling process of  Figure 3 (a), the entanglement entropy increases rapidly and it remains increased after the object tunnels through the barrier (Figure 5 (a)). Contrarily, although the entropy changes slightly during the tunnelling process,  it  remains almost zero in Figure 1 (a) (Figure 5 (b)). In such a situation, the decoherence effect discussed in this section can not be observed.

\begin{figure}
\begin{center}
{%
\includegraphics[clip,width=1\columnwidth]{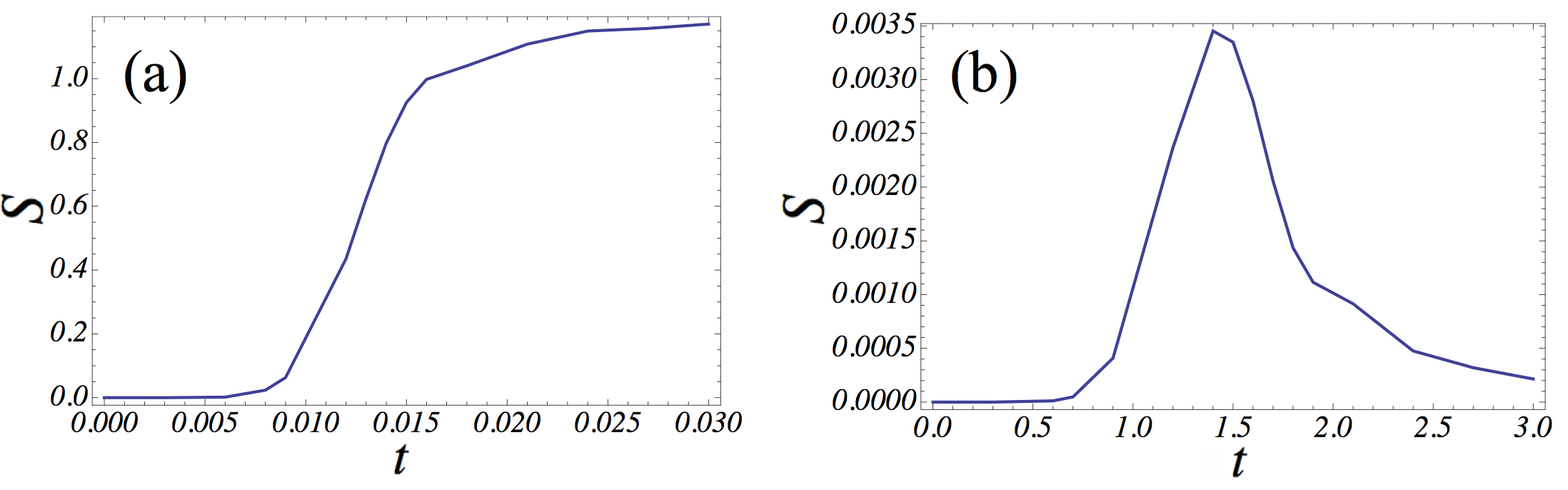}%
}
\caption{The time evolution of the entanglement entropy $S$ between the c.m. and relative coordinate degrees of freedom during the tunnelling process. (a) is for the tunnelling process in the Figure 3 (a), the object starts to tunnel the barrier around $t=0.01$. (b) is for the tunnelling process in the Figure 1 (a), the object starts to tunnel the barrier around $t=1$. }
\end{center}
\end{figure}

\section{The case for gravity}

\subsection{Composite object coupled with the quantized gravitational field}\label{gravity}

In the previous sections, we dealt with the photon radiation field. Because of similarity between linearized gravity and electromagnetism, it is also possible to deal with the gravitational radiation field using the methods presented in the previous sections. 

Quantization of gravity by analogy with the quantization of the electromagnetic field and the development of the path integral formalism in a curved space have been long investigated in the literature \cite{dewitt}. Recently the studies of gravitational decoherence have attracted interest \cite{pik, cisco,penrose,diosi,ble, cisco2, mari}. However, in many theoretical studies, object concerned and its trajectories derived from its Hamiltonian are often not specified in detail. On the other hand, decoherence, even though driven by interaction with an environment, can depend on the details of the object concerned and its time evolution. Here we derive the master equation of the composite object introduced in this article, which is now coupled with the quantized gravitational radiation field by its energy-momentum tensor (i.e., quadrupole moment) in the weak gravity regime. The master equation can be derived by the similar method employed for the photon radiation field. However here we need to introduce the strong assumptions that  an observer is external, is in the source-free region, and the energy-momentum tensor of the object and the gravitational radiation field generated by it can be quantized. Then we show that the spatial superposition state of the object prepared as in Figure \ref{dec} can decohere if ``quantized'' gravitational waves can be traced out.

Let us consider the situation where two constituent particles of the composite object do not have charges, however they are still coupled to the gravitational radiation field $h_{ij}$ by their energy-momentum tensor $T^{ij}$. It is generally assumed that an observer in quantum mechanics is always external and is located in the source-free region. With this assumption, the transverse-traceless gauge, i.e., $h_{0\mu}=h^{ij}_{\mbox{ }, i}=h^{i}_{\mbox{ }i}=0$, can be introduced. Then $S_{\rm int}$ and $S_{\rm rad}$ in (\ref{action}) are replaced by 
\begin{eqnarray}\label{TTgauge}
S_{\rm int}&=&\int d^4 r h_{ij} (\mathbf{r},t)T^{ij} (\mathbf{r},t),\nonumber\\
S_{\rm rad}&=& \frac{1}{4\pi G}\int d^4 r h_{\mu\nu ,\alpha}h^{\mu\nu,\alpha}=\frac{1}{4\pi G}\int d^4 r (\partial_{t}h_{ij}\partial^{t}h^{ij}-h_{ij,k}h^{ij,k})
\end{eqnarray}
in the weak gravity regime \cite{pauli}. Here Latin indices represent spatial dimensions, i.e., $i, j, k=1,2,3$ and Greek indices represent spacetime dimensions, i.e., $\mu,\nu,\alpha, \beta =0,1,2,3$.

As in the case of the photon radiation field, we introduce the Fourier expansion so that the transverse-traceless parts $h_{ij}$ of the gravitational field and $T^{ij}$ can be written as
\begin{eqnarray}
h_{ij} (\mathbf{r}, t) &=&\displaystyle\sum_{\mathbf{k},\lambda=1,2}h_{\mathbf{k}ij}^{(\lambda)}e^{i\mathbf{k}\cdot\mathbf{r}}, \mbox{ } T^{ij} (\mathbf{r},t) =\displaystyle\sum_{\mathbf{k}}T^{ij }(\mathbf{k},t)e^{i\mathbf{k}\cdot \mathbf{r}}
\end{eqnarray}
where $h_{\mathbf{k}ij}^{(\lambda)}=\epsilon_{ij}^{(\lambda)}s_{\mathbf{k}}^{(\lambda)}(t)$ with the transverse-traceless polarization tensors $\epsilon_{ij}^{(\lambda)}$.

The influence functional for coupling to the gravitational radiation field can be obtained similarly to the  case of the photon radiation field \cite{fumika, ana, hu}:
\begin{eqnarray}\label{influence2}
\mathcal{F}(T^{ij},T'^{ij})=\exp i\Theta_{\rm grav}\nonumber\\
=\exp \Biggr[i\displaystyle\sum_{\mathbf{k}}\int_0^{t}\int_0^{s}dsds' (T^{ij}(\mathbf{k},s)-T'^{ij}(\mathbf{k},s))\gamma_{ij,kl}^{(s,s')}\nonumber\\
\qquad\qquad\qquad\qquad\times (T^{*kl}(\mathbf{k},s')+T'^{*kl}(\mathbf{k},s'))\nonumber\\
 -\displaystyle\sum_{\mathbf{k}}\int_0^{t}\int_0^{s}dsds' (T^{ij}(\mathbf{k},s)-T'^{ij}(\mathbf{k},s))\eta_{ij,kl}^{(s,s')}\nonumber\\
 \qquad\qquad\qquad\times (T^{*kl}(\mathbf{k},s')-T'^{*kl}(\mathbf{k},s'))\Biggr]
\end{eqnarray}
with the correlation functions
\begin{eqnarray}
\gamma_{ij,kl}^{(s,s')}&=&2\pi G\frac{ \Pi_{ij,kl} }{k}\sin k (s-s'),\quad \eta_{ij,kl}^{(s,s')}=2\pi G\frac{ \Pi_{ij,kl}}{k}\cos k (s-s')
\end{eqnarray}
where the sum over polarizations of two on-shell graviton states gives the polarization tensor $\displaystyle\sum_{\lambda}^{2} \epsilon^{\lambda}_{ij} (\mathbf{k}) \epsilon^{*\lambda }_{kl} 
(\mathbf{k}) =\Pi_{ij , kl}  = \frac12 (\delta_{ik}\delta_{jl}+\delta_{il}\delta_{jk}-\delta_{ij}\delta_{kl})$. 

Here $T^{ij}$ and $T'^{ij}$ are associated with $\mathbf{x}$ and $\mathbf{x}'$ respectively. If we introduce the energy-momentum tensor,
\begin{eqnarray}
T^{ij} (\mathbf{k},s)=\mu \dot{x}^{i}(s)\dot{x}^{j} (s) e^{i\mathbf{k}\cdot \mathbf{x}(s)}
\end{eqnarray}
where the relative coordinate degrees of freedom $\mathbf{x}$ obeys an equation of motion for a harmonic oscillator (\ref{semiclassical}) as before, then we can write
\begin{eqnarray}\label{Theta2}
&&\mbox{Re}(\Theta_{\rm grav})\approx\frac{G\mu^2\omega}{2\pi}\int_0^{t}ds (\dot{x}(s)^4-\dot{x}'(s)^4)\nonumber\\
&&+\frac{G\mu^2\omega^2}{2}\int_0^{t}ds (\dot{x}(s)^2-\dot{x}'(s)^2)(\dot{x}(s)x(s)+\dot{x}'(s)x'(s)),\nonumber\\
&&\mbox{Im}(\Theta_{\rm grav})\nonumber\\
&&\approx\frac{3G\mu^2 \omega^5 t }{16\pi}(x^2-x'^2)^2\nonumber\\
&&+\frac{G\mu^2\omega^2}{4\pi}\int_0^{t}ds (\dot{x}(s)^2-\dot{x}'(s)^2)(x(s)\dot{x}(s)-x'(s)\dot{x}'(s))
\end{eqnarray}
and  the master equation  can be obtained as
\begin{eqnarray}
\frac{\partial \rho (X,X',x,x',t)}{\partial t}\approx\Biggr[ \frac{i}{2M}\left(\frac{\partial^2}{\partial X^2}-\frac{\partial^2}{\partial X'^2}\right)+\frac{i}{2\mu}\left(\frac{\partial^2}{\partial x^2}-\frac{\partial^2}{\partial x'^2}\right)\nonumber\\
-i (V'(x,X)-V' (x',X'))-\frac{3G\omega i}{2\mu^2\pi}\left(\frac{\partial^2}{\partial x^2}-\frac{\partial^2}{\partial x'^2}\right)\left(\frac{\partial^2}{\partial x^2}+\frac{\partial^2}{\partial x'^2}\right)\nonumber\\
+\frac{G\omega^2}{\mu}\left(x\frac{\partial}{\partial x}-x'\frac{\partial}{\partial x'}\right)\left(\frac{\partial^2}{\partial x^2}-\frac{\partial^2}{\partial x'^2}\right)-\frac{3G\mu^2 \omega^5 }{16\pi}(x^2-x'^2)^2\nonumber\\
-\frac{G\omega^2 i}{4\mu\pi}\left(x\frac{\partial}{\partial x} + x'\frac{\partial}{\partial x'}\right)\left(\frac{\partial^2}{\partial x^2}-\frac{\partial^2}{\partial x'^2}\right) \Biggr]\rho (X,X', x,x',t)
\end{eqnarray}
where $V' (x,X)=\frac12 \mu \omega^2 x^2+V(x,X)$ and $V(x,X)$ is the potential barrier as before.

The effects of dissipation and decoherence caused by tracing out the gravitational radiation field here are similar to those given by the photon radiation field discussed in the previous sections. Emission of gravitational waves causes the decay of the quantum tunneling rate when the relative coordinate degrees of freedom of the object is initially prepared in the excited state. Decoherence of the spatial superposition state of the object occurs if the relative coordinate degrees of freedom gets entangled with c.m. degrees of freedom by propagation through the barrier. However the decay rates are much  smaller than those of the photon counterpart by a factor of $G\mu^2/q^2$  and therefore unobservable on the time scales used in the previous sections. 

Nevertheless, the simple model discussed here can provide the spatial superposition state of the c.m. degrees of freedom which is entangled with the quadrupole moment represented by the oscillator in the relative coordinate of its constituent particles, and therefore has a possibility to exhibit decoherence due to gravitational waves if they can be quantized.

\subsection{Lorenz gauge and false loss of coherence}\label{lorenzgauge}

So far we used the Coulomb gauge or the transverse-traceless gauge assuming that the observer is in the source-free region. Of course one can also use the Lorenz gauge. In the Lorenz gauge, in addition to the contribution of the radiation field (i.e., real photons or gravitons), one may also write the contributions of virtual particles, i.e., the self-interaction terms. Unlike real particles, those virtual particles do not decohere the superposition of the excited states and the ground state of composite objects, rather they can cause false loss of coherence due to their inability to detach themselves from the object concerned that generates them \cite{bill}. As an example, it was discussed in \cite{bill} that the spatial superposition state of an electron with its attached static Coulomb field can decohere due to differences of the states of the  field associated with two different positions of the electron, however if these two positions are made come together by the force field, it can be seen that coherence can be restored  and therefore decoherence caused by the static Coulomb field at the intermediate times is not real. This happens because the static  field mediated by virtual particles can not detach themselves from the object concerned (i.e., electron), and can not carry away the information.

\begin{figure}
\begin{center}
{%
\includegraphics[clip,width=0.7\columnwidth]{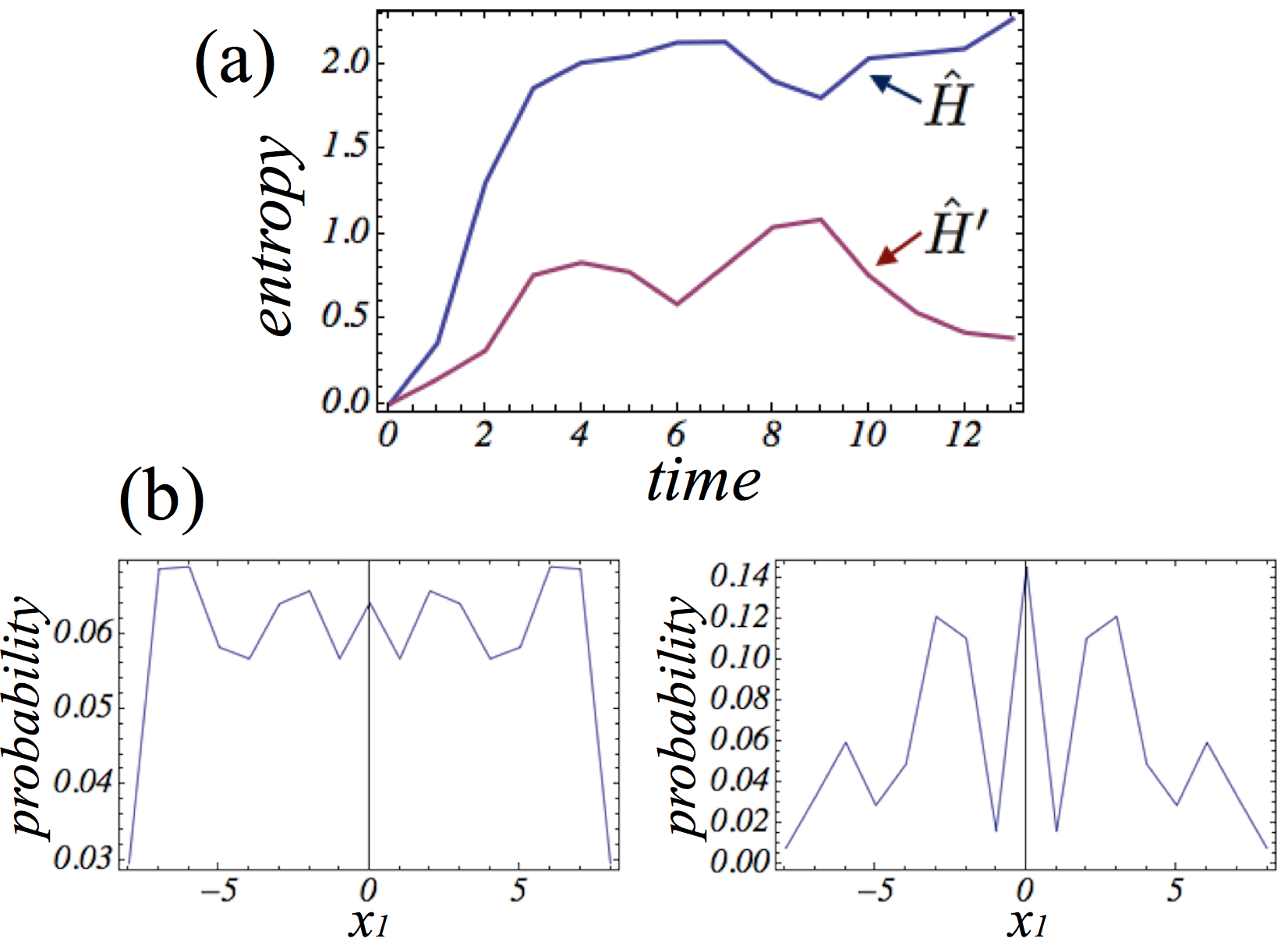}%
}
\caption{(a) The time evolution of the entanglement entropy which measures the entanglement between $x_1$ and $x_2$ given by $\hat{H}$ and $\hat{H}'$ respectively. (b) Interference pattern produced by the reduced density matrix in $x_1$. The time evolution of the density matrix is given by $\hat{H}$ (left) and $\hat{H}'$ (right) respectively.}
\label{restore}
\end{center}
\end{figure}

Similar processes that restore coherence can also occur even when real particles are traced out if they are controlled in a particular manner. For example, let us consider an electron which is in a spatial superposition state of being at $x_1=d>0$ or $x_1=-d$ whose wave-function is written as
\begin{eqnarray}\label{gaussian}
\psi (x_1) =\frac{1}{\sqrt{2}}\frac{1}{(2\pi \sigma^2)^{1/4}}\left(e^{-\frac{1}{4\sigma^2} (x_1-d)^2}+e^{-\frac{1}{4\sigma^2} (x_1 +d)^2}\right)\nonumber\\
\end{eqnarray}
where the width $\sigma$ of the Gaussian distribution is  smaller than $d$ so that two Gaussians have essentially no overlap. Here we choose $d/\sigma =5$.

Instead of the situation discussed above where we only had the electron and the Coulomb field generated by it, we now add a second electron that interacts with the first one through the Coulomb field. Then the Hamiltonian can be written as
\begin{eqnarray}\label{false}
\hat{H}=\frac{\hat{p}_1^2}{2m}+\frac{\hat{p}_2^2}{2m}+\frac{q^2}{4\pi\epsilon_0 |\hat{x}_1-\hat{x}_2|}.
\end{eqnarray}

We place the second electron at the origin $x_2=0$ so that its wave-function can be written as
\begin{eqnarray}
\phi (x_2) =\frac{1}{(2\pi\sigma^2)^{1/4}}e^{-\frac{1}{4\sigma^2}x_2^2}.
\end{eqnarray}

If we start with the initial wave-function  which is in the pure state, $\psi (x_1)\phi(x_2)$, the time evolution by the Hamiltonian (\ref{false}) makes  two electrons get entangled with each other since if the second electron is repelled in the negative direction, we know that the first one is at $x_1=d$, while it is at $x_1=-d$ if the second one is repelled in the positive direction. As a result, tracing out the position $x_2$ of the second electron causes the spatial superposition state (\ref{gaussian}) of the first electron to decohere. Indeed, it can be seen that the  entanglement entropy that measures the entanglement between $x_1$ and $x_2$ increases with time when the whole system evolves according to $\hat{H}$ (Figure \ref{restore} (a)) and the interference pattern produced by the reduced density matrix in $x_1$ (obtained by tracing out $x_2$) when two Gaussians of (\ref{gaussian}) overlap and interfere with each other after the time evolution, is largely suppressed (Figure \ref{restore} (b, left)). 

Now we attach the second electron to a spring:
\begin{eqnarray}
\hat{H}'=\hat{H}+\frac12 c x_2^2
\end{eqnarray}
so that the spring tries to pull it back toward the origin after some time. Then the second electron becomes  bounded in a limited region and one is able to restore coherence. In Figure \ref{restore} (a), it can be seen that the entropy increases initially, but the spring tries to pull the second electron back when it gets far away from the origin and the entropy decreases again if the system evolves according to $\hat{H}'$ and the clear interference pattern (Figure \ref{restore} (b, right)) can be observed. In the figures, $\frac{c\sigma}{q^2 /(4\pi \epsilon_0\sigma^2)}=1/200$ is chosen.  

This type of behavior, which is often called a recurrence phenomenon, can be seen when the system is bounded in a limited region of space \cite{poin, sch}. For example, if we have real photons that are emitted and travel off to infinity so that they can not come back into the system, then the information is carried away by them and coherence can not be restored. On the other hand, if we put mirrors that bounce the photons back into the system, then the information and coherence can be restored.

False loss of coherence \cite{bill} caused by tracing out virtual particles and soft particles that are tied to the system is the extreme case where the information can not even be detached from the system by those particles traced out. Therefore, the Coulomb gauge and the transverse-traceless gauge are generally useful when one is only interested in the far field radiation problems and real loss of coherence. 

The discussions above would be valid even if photons are replaced by gravitons, as long as one introduces the strong assumptions as we have done in section 3.1. However, in general relativity, the observer is normally considered as internal, and is affected by spacetime structure made by the system concerned and himself and his backreaction on it etc. Furthermore, matter causes the flow of time to vary. It was then suggested in \cite{penrose, diosi} that the superposition of matter causes the superposition of time  which may lead to gravitational decoherence as the quantum gravity effect.  It is still an open problem how the quantum mechanics of an internal observer \cite{aha2, aha3} or quantum reference frame \cite{bill3,pope} can be related to those studies of gravitational decoherence, and the relations between observables of the internal observer and those of external observer or those evaluated by one coordinate/gauge fixing and by the other coordinate fixings \cite{cisco2} are still unclear. Those questions are beyond the scope of section 3.1, and we performed all calculations of section \ref{gravity} with the assumption that the observer is external and is in the source-free region in flat spacetime.

\section{Conclusion}

Unlike a single point particle coupled to the environment, a composite object coupled to the environment can exhibit complicated quantum behavior. It is often the case that the environment acts only on one degrees of freedom of the object, while it has observable consequences for the other degrees of freedom of it, since these degrees of freedom of the object can be entangled or correlated with each other. As a relatively simple example, we studied  quantum tunneling of a composite object,  whose dipole or quadrupole moment is coupled to the photon or gravitational radiation field, through a $\delta$ potential barrier.  

Many studies have been done on dissipation and decoherence effects in quantum mechanics of a single particle coupled to the environment.  Meanwhile, when one studies decoherence caused by dipole and quadrupole moment radiation, normally decoherence acts on superposition states involving two or more particles consisting the corresponding dipole and quadrupole moment.  In \cite{bre, bre2, fumika}, dipole moment or quadrupole moment  of the composite system consisting of a particle and device of  superposition experiment (e.g., slit, mirror) was considered.  In such a situation, not only a particle, but the device is also considered to be in the superposition state, i.e., $|$particle passing through left slit$>$$|$left slit kicks the particle to the right$>+$$|$particle passing through right slit$>$$|$right slit kicks the particle to the left$>$ where the kick gives the momentum transfer from the device (slit) to the particle so that superposed paths get recombined. In this way, superposition of dipole moment or quadrupole moment of the composite system (particle+device) was introduced in \cite{bre, bre2, fumika}. 
However, in a realistic situation, it may be easier to  access parameters involving superposition states of atom or molecule with its  dipole or quadrupole moment. This is another reason why we attempted to study quantum mechanics of composite object (instead of a single particle) coupled with  quantized radiation field in this article.

Firstly, it was observed that although the c.m. degrees of freedom of the object does not directly interact with the radiation field, dissipation caused by it can decrease the quantum tunneling rate of the object in the c.m. degrees of freedom. This phenomenon can be seen when the initial state of the oscillator in the relative coordinate is prepared in the excited state. It is because the state  can decay into the ground state due to the photon emissions, and the object with its oscillator being in the ground state sees the high potential barrier.

Secondly,  we have seen that the spatial superposition state of the object created by the tunnelling process can decohere by the coupling between its relative coordinate degrees of freedom and the radiation field. We found that the propagation of the object through the barrier can make its c.m. and its relative coordinate degrees of freedom entangled with each other when the initial energy of the c.m. motion is similar to that of the excited states of the  oscillator. Consequently, tracing out the radiation field results in suppression of the spatial superposition state of the object in the c.m. degrees of freedom. On the other hand, this fact demonstrates that one is able to avoid such decoherence by preventing the creation of the entanglement between  the c.m and the relative coordinate degrees of freedom while the object interacts with the barrier, e.g., by decreasing the energy transfer between the two degrees of freedom. Note that, in this article, we had the relatively simple entangled state created which is similar to the GHZ state,  $\approx |\mbox{c.m. transmit}\rangle|\mbox{rel ground state}\rangle |\mbox{no photon emission}\rangle +|\mbox{c.m. reflected}\rangle|\mbox{rel excited states}\rangle|\mbox{photon emissions}\rangle+\cdots$ up to normalization. With more complicated entangled states, the notion of the superposition state and decoherence can become more involved.  

The work shows that the study of quantum behavior of a composite object coupled to the external environment requires careful handling of different types of interplays, (i) among the c.m, the relative coordinate degrees of freedom and the potential barrier, and (ii) between the relative coordinate degrees of freedom and the radiation field (i.e., environment), and it can provide rich complex phenomena which can not be seen in quantum physics of a single point particle coupled to the environment.

Finally, due to the increasing interest in gravitational decoherence, we discussed the situation in which the photon field is replaced by gravity. Since linearized gravity is similar to electromagnetism,  it is possible to derive the master equation in a similar way to that for photons, but with the strong assumptions that that   observer is external, is in the source-free region, and the energy-momentum tensor of the object and the gravitational radiation field generated by it can be quantized. We concluded that our simple model can be used to exhibit the decoherence due to gravitational waves if they can be quantized since it provides the spatial superposition state of the object which is entangled with the superposed quadrupole moment.

\ack
This work is largely benefited from fruitful discussions with W. G. Unruh and we appreciate his important comments. We also thank B. L. Hu, G. W. Semenoff, M. Hotta, J. J. Halliwell, C. Anastopoulos, T. Momose and Y. Shikano for helpful discussions and support. F. S. is partially supported by a UBC international tuition award, UBC faculty of science graduate award, the research foundation for opto-science and technology and  DAAD Research Grants. The work was supported by NSERC Discovery grant and CFI funds for CRUCS.

\end{document}